\documentclass[apj,onecolumn]{emulateapj}

\usepackage{amsmath}
\usepackage[normalem]{ulem}
\usepackage{color}

\newcommand{\be}{\begin{equation}}
\newcommand{\ee}{\end{equation}}
\def\lesssim{\buildrel < \over {_{\sim}}}
\def\gtrsim{\buildrel > \over {_{\sim}}}

\shorttitle{Collisionless shock in a partially ionized medium:\\
III. Efficient cosmic ray acceleration}
\shortauthors{Morlino et al.}

\begin{document} 

\title{Collisionless shocks in a partially ionized medium: III. 
\\ Efficient cosmic ray acceleration}

\author{G. Morlino\altaffilmark{1}, P. Blasi\altaffilmark{1,2}, 
R. Bandiera\altaffilmark{1}, E. Amato\altaffilmark{1}, 
D. Caprioli\altaffilmark{3}}
\affil{$^{1}$INAF-Osservatorio Astrofisico di Arcetri, Largo E. Fermi, 5, 50125
Firenze, Italy}
\affil{$^{2}$INFN/Laboratori Nazionali del Gran Sasso, Assergi, Italy}
\affil{$^{3}$Department of Astrophysical Sciences, Peyton Hall, Princeton
University, Princeton, NJ 08544, USA}

\begin{abstract}
In this paper we present the first formulation of the theory of non-linear
particle acceleration in collisionless shocks in the presence of neutral
hydrogen in the acceleration region. The dynamical reaction of the accelerated
particles, the magnetic field amplification and the magnetic dynamical effects
on the shock are also included. The main new aspect consists however in
accounting for charge exchange and ionization of neutral hydrogen, which
profoundly change the structure of the shock, as discussed in our previous work.
This important dynamical effect of neutrals is mainly associated to the
so-called neutral return flux, namely the return of hot neutrals from the
downstream region to the upstream, where they deposit energy and momentum
through charge exchange and ionization.
We also present the self-consistent calculation of Balmer line emission from the
shock region and discuss how to use measurements of the anomalous width of the
different components of the Balmer line to infer the cosmic ray acceleration
efficiency in supernova remnants showing Balmer emission: the broad Balmer line,
which is due to charge exchange of hydrogen atoms with hot ions downstream of
the shock, is shown to become narrower as a result of the energy drainage into
cosmic rays, while the narrow Balmer line, due to charge exchange in the
cosmic-ray--induced precursor, is shown to become broader.
In addition to these two well-known components, the neutral return flux leads to
the formation of a third component with intermediate width: this too contains
information on ongoing processes at the shock.
\end{abstract}


\section{Introduction} \label{sec:intro}

In the context of the  so-called Supernova Remnant (SNR) paradigm for the
origin of Galactic Cosmic Rays (CRs), the bulk of CRs are accelerated at SNR
shocks through diffusive shock acceleration (DSA). Under this assumption, in
order to explain the observed fluxes at Earth and, at the same time, the B/C ratio
(sensitive to the grammage traversed by CRs during propagation) one is forced to
require an acceleration efficiency of $5-15\%$ per SNR, depending on details of
the propagation in the Galaxy \cite[e.g.][]{amato1,amato2}. In these conditions,
the non-linear effects induced by the dynamical reaction of accelerated
particles on the SNR shock cannot be neglected and reflect on the shape of the
spectrum \cite[e.g.][for a review]{jones91, maldrury} and on the temperature of the ionized plasma
downstream of the shock \cite[]{decourchelle+00, drury09, patnaude09, castro11}. 
Moreover, the streaming instability induced by
accelerated particles ahead of the shock is thought to be responsible for the
magnetic field amplification that is a crucial ingredient to reach maximum
energies close to the knee \cite[]{dam2007} and to explain the thin non-thermal
filaments observed in X-rays \cite[e.g.][]{rims}. 

The non-linear theory of diffusive
shock acceleration (NLDSA) has been developed by several authors 
through several different approaches: two-fluid models \cite[e.g.][]{drury81}, 
kinetic models \cite[e.g.][]{malkov97, maldia00,Blasi02}
and numerical approaches, both Monte Carlo and other 
simulation procedures \cite[e.g.][]{don90, don95, don96, kang02, kang05, vladi06}.

The description of modified shocks we adopt in the following, however, is based on 
a semi-analytical model presented in a series of papers by our group
\cite[]{AmatoBlasi05,AmatoBlasi06,Caprioli09}. For fully ionized plasmas, this model
gives a complete description of the full non-linear chain of processes related to
efficient particle acceleration, including magnetic field amplification due to 
streaming instabilities, and the backreaction of the 
amplified field on the system. In terms of amplifying the field, both resonant 
\cite[e.g.][]{bell78} and non-resonant modes \cite[]{bell04} of the streaming 
instability can contribute. 
However the non-resonant mode is typically important
only in the case of very fast shocks \cite[]{AmatoBlasi09}, corresponding to 
few hundred years old remnants, for a typical evolution scenario. Moreover, while 
the magnetic fields inferred in several SNRs seem to be in agreement with the saturation levels predicted by 
\cite{bell04} \cite[e.g.][]{volk05}, it is not clear at all that the non resonant mode can enhance particle
scattering, which is what is required for more efficient acceleration: this is
because this mode has very short wavelengths compared with the typical particles'
gyroradius, and very efficient inverse cascading is needed before it can resonantly 
scatter the particles.

Both test-particle DSA and NLDSA
lead to predict spectra which are rather hard at high energies, with a power-law
index in momentum $\lesssim 4$. This implies a steep dependence on energy of the
Galactic diffusion coefficient \cite[]{berevolk2007}, if the CR spectrum at
Earth is to be reproduced. Such energy dependence appears to be at odds with the
observed anisotropy of Galactic CRs \cite[]{ptuskin06,amato2}. Accounting for
the velocity of scattering centers in the amplified magnetic field
might lead to somewhat steeper injection
spectra thereby mitigating the problem \cite[]{damiano}.

In general, anything that affects the structure of the shock also leads to
changes in the spectrum of accelerated particles. One important source of shock
modification is represented by the presence of neutral atoms in the acceleration
region: as discussed by \citet[hereafter Paper I]{paperI}, the reactions of
charge exchange (CE) and ionization produce important effects in the shock
region. The most important of these effects is induced by the so-called {\it
neutral return flux} (NRF): a hydrogen atom crossing a collisionless shock front
can undergo a charge-exchange with a hot ion downstream of the shock, thereby leading to
the formation of a cold ion and a hot neutral. It may happen that the latter is
produced with velocity in the direction of the shock; being neutral, it can
recross the shock and suffer CE or ionization upstream, hence releasing energy
and momentum there. 

The existence of a NRF was highlighted for
the first time by \citet{limraga}. These authors, however, adopted a very low fraction
of neutrals, so that the effect of the NRF on the system dynamics was not very important
(in addition the neutrals were treated as maxwellian, which is not a good approximation and
may lead to an incorrect estimate of the NRF). The dynamical role of the returning neutrals
was worked out  under general conditions in Paper I, where we
showed how these lead to heating of the upstream
ionized plasma and to a decrease of the Mach number. Even in the context of test-particle 
DSA it is easy to show that this phenomenon leads to a steepening of
the spectrum of particles accelerated in a SNR where appreciable neutral
hydrogen is present (Paper I). The NRF affects the upstream plasma on a scale of
a few interaction lengths of CE. 

The presence of neutral hydrogen also represents a unique diagnostic tool of the
acceleration process, through the Balmer emission associated with CE to excited
states: in a Balmer dominated shock \cite[]{raymond78}, the Balmer
line has a narrow and a broad component \cite[see][for a review]{balmer},
the first associated with hydrogen atoms that
did not suffer CE with higher temperature ions, and the second that reflects the
temperature of shocked ions, due to CE reactions downstream
\cite[]{vanadel,cheva80}. In the presence of CR acceleration the ions'
temperature downstream is lower, and this should reflect in decreased Full-Width at 
Half-Maximum (FWHM) of the broad Balmer line. 
If the CR pressure is large enough, a CR-induced precursor is
formed in front of the shock \cite[]{maldrury} so that CE can also take place
upstream and eventually lead to a broadening of the narrow Balmer line. In
addition, it was shown in \citet[hereafter Paper II]{paperII}, that the shape of
the Balmer line is also affected by the NRF discussed above through the
formation of an intermediate component of the line. 

Remarkably, anomalous widths of both broad and narrow lines have been observed
in Balmer-dominated shocks. For instance, \cite{Helder09} combined proper-motion
measurements of the shock and broad H$\alpha$ line width for the remnant RCW 86
to demonstrate that the inferred low temperature behind the shock may be
interpreted as a signal that a sizable fraction of the energy is being channeled
into CRs. A similar conclusion has been reached by \cite{Helder10} by analizing
the SNR 0509-67.5 in the LMC, even if in this case the shock velocity has been
estimated using an evolution model with information derived from the line width
of elements in the shocked ejecta (wich can be used to infer the speed of
reverse shock).

Anomalous width of narrow Balmer lines have been also observed in several SNRs
\cite[see, e.g.][]{Sollerman03}. The width of such lines is in the 30--50 km
s$^{-1}$ range, implying a pre-shock temperature around 25,000--50,000 K. If
this were the ISM equilibrium temperature there would be no atomic hydrogen,
implying that the pre-shock hydrogen is heated by some form of shock precursor
in a region that is sufficiently thin so as to make collisional ionization
equilibrium before the shock unfeasible. The CR precursor is the most plausible
candidate to explain such a broadening of the narrow line width.

A previous attempt to include neutral particles in the shock acceleration theory
has been done by \cite{Morlino10} using a fluid approach. Nevertheless, as
discussed in Paper I, neutrals cannot be treated as a fluid because the length scales
involved are much smaller than the equilibration scale.
More recently \cite{raymond2011} tried to describe the interaction of neutrals
with the CR precursor using a phenomenological approach where both the
CR spectrum and spatial distribution are assumed rather than calculated.

An accurate description of the CR acceleration process in the presence of
neutrals can only be achieved by formulating a NLDSA theory where the dynamical
action of neutrals is taken into account. This is a needed preliminary step, if
one wants to make use of measurements of  Balmer line emission from SNRs to
quantitatively assess their efficiency as particle accelerators. 
In this paper we present the first treatment of NLDSA in partially ionized media
and discuss how to use the theory to  calculate the shape of the Balmer line and
infer information on particle acceleration at the shock. 

The paper is structured as follows: in Section~\ref{sec:calc} we describe the
calculations used to describe neutrals, ions and the acceleration
process and we explain in detail how to assemble the parts together to describe
NLDSA in the presence of neutrals. The formalism previously introduced in Paper
I --- as far as neutrals are concerned --- and in \cite[]{AmatoBlasi05,
AmatoBlasi06, Caprioli09} --- as far as NLDSA is concerned --- will be heavily
used. In Section~\ref{sec:results} we illustrate our results in terms of shock
modification, spectrum of accelerated particles and shape of the Balmer line. We
summarize and conclude in Section~\ref{sec:conclusion}.

\section{Calculations} \label{sec:calc}

We consider a plane-parallel shock wave propagating in a partially ionized
proton-electron plasma with velocity $V_{\rm sh}$ along the $z$ direction. The
fraction of neutral hydrogen is fixed at upstream infinity where ions and
neutrals are assumed to be in thermal equilibrium with each other. The shock
structure is determined by the interaction of CRs and neutrals with the
background plasma. Both CRs and neutrals profoundly change the shock structure,
especially upstream where both create a precursor: the CR-induced precursor
reflects the diffusion properties of accelerated particles and has a typical
spatial scale of the order of the diffusion length of the highest energy
particles. The neutral-induced precursor develops on a spatial scale comparable
with a few interaction lengths of the dominant process between CE and
ionization. The downstream region is also affected by the presence of both CRs
and neutrals and the velocity gradients that arise from ionization have a direct
influence on the spectrum of accelerated particles.

A self consistent description of the overall system requires to consider four
mutually interacting components: thermal particles (protons and electrons), neutrals
(hydrogen), accelerated protons (CRs) and turbulent magnetic field. We neglect
the presence of helium and heavier chemical elements. In the following
subsections we write down the basic equations used to describe each component,
including the interaction terms. We are looking for  stationary solutions, hence
all equations are written as time-independent. The interaction terms make the
system highly non linear and the solution is found using a iterative scheme
similar to the one we introduced in some of our previous work \cite[][Paper
I]{Blasi02,AmatoBlasi05}. The details on how to generalize such scheme to the
case in which neutrals are present are discussed in \S~\ref{sec:algorithm}.

\subsection{Neutral distribution}
\label{sec:neutrals}

The behavior of neutrals has been extensively discussed in Paper I. Neutral
hydrogen interacts with protons through CE and ionization and with electrons
through ionization only. The hydrogen distribution function, $f_{N}(\vec v,z)$,
can be described using the stationary Boltzmann equation
\begin{equation} \label{eq:vlasov}
v_z \frac{\partial f_{N}(\vec v, z)}{\partial z} = \beta_{N} f_{i}(\vec v, z)  -
        \left[ \beta_{i} + \beta_e \right] f_{N}(\vec v, z) \,,
\end{equation}
where $z$ is the distance from the shock (which is located at the origin),
$v_z$ is the velocity component along the $z$ axis and the electron and proton
distribution functions, $f_i(\vec v,z)$ and $f_e(\vec v,z)$, are assumed to be
Maxwellian at each position. The collision terms, $\beta_k f_l$, describe the
interaction (due to CE and/or ionization) between the species $k$ and $l$. The
interaction rate $\beta_k$ is formally written as
\begin{equation} \label{eq:beta_k}
\beta_k (\vec v,z) = \int d^{3} w \, v_{rel} \, \sigma(\vec v_{rel})
                  f_{k}(\vec w,z) \,,
\end{equation}
where $v_{rel} = |\vec v- \vec w|$ and $\sigma$ is the cross section for the
relevant interaction process. More precisely, $\beta_N$ is the rate of CE of an
ion that becomes a neutral, $\beta_i$ is the rate of CE plus ionization of a
neutral due to collisions with protons, while $\beta_e$ is the ionization rate
of neutrals due to collisions with electrons. A full description of the cross
sections used in the calculations can be found in Paper~II.

A novel method to solve Eq.~(\ref{eq:vlasov}) has been developed in Paper I,
where we showed how to calculate $f_{N}$ starting from the distribution of
protons and electrons. The method is based on decomposing the distribution
function $f_{N}$ in the sum of the neutrals that have suffered 0, 1, 2, ... ,
$k$ processes of charge exchange. Each distribution function is named
$f_{N}^{(k)}$, and clearly $f_{N} = \sum_{k=0}^{\infty} f_{N}^{(k)}$. Using this
formalism, Eq.~(\ref{eq:vlasov}) can be rewritten as a set of $k+1$ equations,
one for each component:
\begin{equation} \label{eq:vlasov_0}
v_z \partial_z f_{N}^{(0)} = -\left[ \beta_{i} + \beta_e \right] f_{N}^{(0)} 
     \quad {\rm for} \; k=0 \,, \quad {\rm and}
\end{equation}
\begin{equation} \label{eq:vlasov_k}
v_z \partial_z f_{N}^{(k)} = \beta_{N}^{(k-1)}
     f_{i} - \left[ \beta_{i} + \beta_e \right] f_{N}^{(k)} 
     \quad {\rm for} \; k>0 \,.
\end{equation}
Each one of these equations is linear, and the solution for $f_{N}^{(k)}$ can
be obtained from $f_{N}^{(k-1)}$. The boundary conditions are imposed at
upstream infinity where ions and neutrals are assumed to start with the same
bulk velocity and temperature, therefore charge exchange occurs at equilibrium
and the distributions do not change. As a consequence, the boundary conditions
are $f_{N}^{(0)}(\vec v, -\infty) = (n_{N,0}/n_{i,0}) f_i(\vec v, -\infty)$ and
$f_{N}^{(k>0)}(\vec v, -\infty)= 0$. A good approximation to the total
distribution function of neutrals is obtained when a sufficient number of
$f_{N}^{(k)}$ is taken into account. This number is determined by the physical
scale of the problem. A rough estimate can be obtained comparing the lengthscale
for CE whith the maximum between the ionization lengthscale and the precursor
length. In general, for reasonable values of the ambient parameters, the longest
scale, and the one relevant for comparison, is that of the CR-induced precursor.

\subsection{CR distribution}
\label{sec:CRspectrum}

The isotropic distribution function of CRs satisfies the following transport equation in
the reference frame of the shock:
\begin{equation} \label{eq:trasp_CR}
 \frac{\partial}{\partial z} \left[ D(z,p) \frac{\partial f}{\partial z}
 \right]
 - u \frac{\partial f}{\partial z} 
 + \frac{1}{3} \frac{d u}{d z} \, p \frac{\partial f}{\partial p} 
 + Q(z,p) = 0  \,.
\end{equation}
The $z$-axis is oriented from upstream infinity $(z=-\infty)$ to downstream
infinity $(z=+\infty)$ with the shock located at $z=0$. We assume that the
injection occurs only at the shock position and is monoenergetic at $p=p_{\rm
inj}$, hence the injection term can be written as $Q(z,p)= Q_0(p) \delta(z)$
where $Q_0(p)= \
({\eta_{\rm inj} n_1}/{4 \pi p_{\rm inj}^2}) \delta(p-p_{\rm inj})$. 
Here $n_1$ is the number density of ions immediately upstream of the subshock, and $\eta_{\rm
inj}$ is the fraction of particles that is going to take part in the
acceleration process. Following \cite{BGV05}, 
$\eta_{\rm inj}$ can be related to the subshock compression factor as:
\begin{equation}
 \eta_{\rm inj} = 4/ \left(3 \sqrt{\pi} \right) \left(R_{\rm sub}-1 \right)
             \xi_{\rm inj}^3 e^{-\xi_{\rm inj}^2} \, .
\end{equation}
Here $\xi_{\rm inj}$ is defined by the relation $p_{\rm inj} = \xi_{\rm inj}
p_{th,2}$ , where $p_{th,2}$ is the momentum of the thermal particles
downstream. $\xi_{\rm inj}$ parametrizes the poorly known microphysics of the
injection process and is taken as a free parameter with a typical value between
2 and 4.

The diffusion properties of particles are described by the diffusion
coefficient $D(z,p)$. We assume Bohm diffusion in the local amplified magnetic
field:
\begin{equation} \label{eq:Diff}
 D(z,p) = \frac{1}{3} c r_L[\delta B(z)] \, ,
\end{equation}
where $r_L(\delta B)= pc/[e \delta B(z)]$ is the Larmor radius in the amplified
magnetic field. The calculation of $\delta B$ is described in
\S~\ref{sec:waves}.

The solution of Eq.~(\ref{eq:trasp_CR}) can be easily obtained by using
standard techniques. Here we generalize the approach of \cite{Blasi02} and
\cite{AmatoBlasi05} in order to relax the assumption of null gradient of the
distribution function downstream (homogeneity), which does not apply to the
general case in which CRs and neutrals are present. 

The solution of Eq.~(\ref{eq:trasp_CR}) in the upstream is the same found by
\cite{AmatoBlasi05}. \cite{boundary} showed that an excellent 
approximation to this solution is:
\begin{equation} \label{eq:f_CR1}
 f_{up}(z,p) = f_0(p) \exp \left[ -\int_z^0 dz' \frac{u(z')}{D(z',p)}  
                           \right] \, ,
\end{equation}
where $f_0(p)= f(0,p)$ is the distribution function at the shock position.
The fact that the expression in Eq.~\ref{eq:f_CR1} provides 
a very good approximation to the solution of Eq.~\ref{eq:trasp_CR} is 
easy to understand: Eq.~\ref{eq:f_CR1} is the exact solution of 
Eq.~\ref{eq:trasp_CR} when the third term can be neglected.
This condition is fulfilled almost everywhere, since the gradient in the
precursor is very localized (at a distance determined by the diffusion
length of the particles that carry most of the energy).

Through the same approach used by \cite{AmatoBlasi05} we can express the
solution in the downstream in the following implicit form:
\begin{equation} \label{eq:f_CR2a}
 f_{down}(z,p) = f_0(p) 
       + \frac{1}{3} \int_0^z \frac{dz'}{D} \int_{z'}^{\infty} dz''
         \frac{du}{dz''} \, p \frac{\partial f}{\partial p} 
         \exp{\left[ -\int_{z'}^{z''} \frac{u}{D} dy \right] } \, .
\end{equation}
We found that an excellent approximation to the expression in
Eq.~(\ref{eq:f_CR2a}) is given by:  
\begin{equation} \label{eq:f_CR2b}
 f_{down}(z,p) = f_0(p) \exp \left\{ -\frac{q_0(p)}{3} \int_0^z \frac{dz'}{D}
          \int_{z'}^{\infty} dz'' \frac{du}{dz''}
               \exp \left[-\int_{z'}^{z''} \frac{u}{D} dz'' \right]     
          \right\} \, ,
\end{equation}
where $q_0(p)= -{d \ln f_0(p)}/{d \ln p}$ is the slope of the distribution
function at the shock position. Both Equations~(\ref{eq:f_CR2a}) and
(\ref{eq:f_CR2b}) give $f(z,p)= f_0(p)$ when the downstream speed $u(z)$ is
constant.

The distribution function at the shock is obtained, as usual, by integrating
Eq.~(\ref{eq:trasp_CR}) across the subshock:
\begin{equation} \label{eq:f_CR0_1}
 \frac{u_1-u_2}{3} p \frac{\partial f_0}{\partial p}
 = \left[ D \frac{\partial f}{\partial z}\right]_1
 - \left[ D \frac{\partial f}{\partial z}\right]_2
 + Q_0(p) \,.
\end{equation}
Here and in the rest of the paper, labels  ``1'' and ``2'' refer to quantities
immediately upstream and immediately downstream of the subshock, respectively.
The value of $\left[ D\,{\partial f}/{\partial z}\right]_1$ can be obtained by
integrating Eq.~(\ref{eq:trasp_CR}) between upstream infinity ($z=-\infty$) and
position 1 \cite[]{Blasi02}. In the standard stationary solution the downstream
distribution is homogeneous, hence $\left[ D\,{\partial f}/{\partial z}\right]_2
\equiv 0$. This condition does not hold in our case because of ionization of
neutrals downstream. The general expression can be derived from
Eq.~(\ref{eq:f_CR2a}) and reads
\begin{equation}
 \left[ D \frac{\partial f}{\partial z}\right]_2
 = -f_0(p) \frac{q_0(p)}{3} \int_0^{\infty} dz \frac{du}{dz'} 
    \exp \left[-\int_0^{z'} \frac{u}{D} dz'' \right] \, .
\end{equation}

Substituting the last expression in Eq.~(\ref{eq:f_CR0_1}) we obtain for
$f_0(p)$:
\begin{equation} \label{eq:sol_f0}
 f_0(p) = \frac{\eta n_0}{4\pi p_{\rm inj}^3} 
  \frac{3 R_{\rm tot}}{R_{\rm tot} U_p(p)-1}
  \exp \left\{ -\int_{p_{\rm inj}}^p \frac{dp'}{p'} 
   \frac{3 R_{\rm tot}}{R_{\rm tot} U_p(p')-1} 
   \left[ U_p(p') + \frac{q_0(p')}{3} K(p') \right]
  \right\}\ ,
\end{equation}
where we defined $U_p(p)= u_p/V_{\rm sh}$ and $u_p(p)$ can be interpreted as an
estimate of the mean plasma velocity that a particle with momentum $p$
experiences in the precursor region \cite[see Equation~(8) in][]{AmatoBlasi05}.
Eq.~(\ref{eq:sol_f0}) differs from the standard solution only for the term
including $K(p)$, where
\begin{equation}
 K(p) = \frac{1}{V_{\rm sh}} \int_0^\infty dz \frac{du}{dz} 
        \exp \left[ -\int_0^z \frac{u}{D} dz' \right] \,.
\end{equation}
The mean plasma speed experienced in the downstream by particles that are able
to return to the subshock is $u_2 + V_{\rm sh}K(p)$. When the downstream plasma
has a constant velocity, $K(p) \rightarrow 0$ and Eq.~(\ref{eq:sol_f0}) reduces
to the standard solution \cite[see Equation~(7) in][]{AmatoBlasi05}.

\subsection{Transport equation for waves}
\label{sec:waves}

Following \cite{AmatoBlasi06} we describe the scattering of accelerated
particles in the acceleration region as due to the waves that the particles
generate through resonant streaming instability. These waves are also damped due
to several processes. In particular, when the plasma is not fully ionized, the
presence of neutrals can damp Alfv\`en waves via ion-neutral damping.  

The equation for transport of waves can be written as:
\begin{equation} \label{eq:wave_tr}
 \frac{\partial F_w(k,z)}{\partial z} = 
    u(z) \frac{\partial P_w(k,z)}{\partial z} + P_w(k,z) \left[ \sigma_{\rm CR}(k,z)
     - \Gamma_{\rm TH}(k,z)   \right] \,,
\end{equation}
where $F_w$ and $P_w$ are, respectively, the energy flux and the pressure per
unit logarithmic bandwidth of waves with wavenumber $k$. $\sigma$ is the growth
rate of magnetic turbulence, while $\Gamma_{\rm TH}$ is the damping rate.  For
resonant wave amplification the growth rate of Alfv\'en waves is
\cite[]{bell78}:
\begin{equation} \label{eq:sigma_CR}
 \sigma_{\rm CR}(k,x)= \frac{4\pi}{3} \frac{v_A(x)}{P_w(k,x)} \left[ 
    p^4 v(p) \frac{\partial f}{\partial x} \right]_{p=\bar p(k)} \,,
\end{equation}
where $p=\bar p(k)= eB/k m_p c$ is the resonant momentum. The damping of the
waves is mainly due to non-linear Landau damping and ion-neutral damping. For
the sake of simplicity here we adopt a phenomenological approach in which
the damping results in a generic Alfv\'en heating at a rate  $\Gamma_{\rm TH} =
\eta_{\rm TH} \sigma_{\rm CR}$. This expression assumes that a fraction
$\eta_{\rm TH}$ of the power in amplified waves is locally damped and results in
heating of the background plasma \cite[]{ellison}. 

Following \cite{Caprioli09}, Eq.~(\ref{eq:wave_tr}) can be solved by
integrating in $k$-space and using the relation between energy density and
pressure of Alfv\`en waves, namely $F_w = 3 P_w u$, which holds upstream when
$v_A \ll V_{\rm sh}$. This leads to the following equation:
\begin{equation} \label{eq:wave_tr2}
 2 U(z) \frac{d \tilde P_w(z)}{dz} = V_A(z) \left[1-\eta_{\rm TH}\right] 
    \frac{d \tilde P_{\rm CR}(z)}{dz}  - 3 \tilde P_w(z) \frac{dU(z)}{dz}  \,,
\end{equation}
where we used the normalized quantities $U=u/V_{\rm sh}$, $V_A=v_a/V_{\rm sh}$,
$\tilde P = P/(\rho_0 V_{\rm sh}^2)$.  Eq.~(\ref{eq:wave_tr2}) can be used to
derive $P_w$ in the upstream region. 

The jump conditions for the magnetic field are calculated following
\cite{Caprioli09}:
\begin{equation} \label{eq:P_w2}
 P_{w,2} = R_{\rm sub} P_{w,1} \,,              
\end{equation}
\begin{equation} \label{eq:F_w2}
 F_{w,2} - F_{w,1} = 2(R_{\rm sub}-1) u_{i,1} P_{w,1} \,. 
\end{equation}
In the downstream section we solve the same transport Eq.~(\ref{eq:wave_tr})
but we assume the relation between energy flux and pressure that derives from
Eqs.~(\ref{eq:P_w2}) and (\ref{eq:F_w2}), namely $F_w= (1+2 R_{\rm sub}) u P_w$.
Moreover, in the downstream region we assume for simplicity that damping occurs
on scales much larger than the  typical scale length for CE and ionization (this
assumption is not strictly needed, it simply makes the results of
the calculations easier to interpret).

\subsection{Dynamics of the ionized gas} 
\label{sec:dynamics}

The dynamics of the background plasma is affected by the presence of
accelerated particles and by CE and ionization of neutrals. Protons and
electrons in the plasma are assumed to share the same local density,
$\rho_i(z)=\rho_e(z)$, but not necessarily the same temperature, i.e., $T_{i}(z)$ may
be different from $T_{e}(z)$. The equations describing the conservation of
mass, momentum and energy taking into account the interactions of the plasma
fluid with CRs are:

\begin{equation} \label{eq:rh1}
 \frac{\partial}{\partial z} \left[\rho_i u_{i} + \mu_N  \right]=0 \,,
\end{equation}
\begin{equation} \label{eq:rh2} 
 \frac{\partial}{\partial z} \left[ \rho_i u_{i}^{2} + P_{g} + P_{c} + P_{w} 
        + P_{N}  \right]=0 \,,
\end{equation}
\begin{equation} \label{eq:rh3}
 \frac{\partial}{\partial z} \left[ \frac{1}{2} \rho_i u_{i}^{3} + 
  \frac{\gamma_{g} P_{g} u_{i}}{\gamma_{g}-1} + F_w + F_{N} \right]
  = -u_i \frac{\partial P_c}{\partial z} + \Gamma P_w \,.
\end{equation}
Here $\mu_N = m_H \int d^{3} v v_{\parallel} f_{N}$, $P_N = m_H \int d^{3} v
v_{\parallel}^{2} f_{N}$ and $F_N = m_H/2 \int d^{3} v v_{\parallel}
(v_{\parallel}^{2} + v_{\perp}^{2}) f_{N}$ are respectively the fluxes of mass,
momentum and energy of neutrals along the $z$ direction (labelled as
$\parallel$).  They can be easily computed once the neutral distribution
function is known. $P_w$ and $F_w$ are the pressure and energy flux of waves as
defined in section \S~\ref{sec:waves}, while $P_c$ is the CR pressure computed
from the CR distribution function:
\begin{equation}
 P_c(z) = \frac{4 \pi}{3} \int dp \, p^3 v(p) f(z,p) \,.
\end{equation}
The dynamical role of electrons in the conservation equations is usually
neglected due to their small mass. However, collective plasma processes could
contribute to equilibrate electron and proton temperatures, at least partially.
If the equilibration occurs in a very efficient manner, the electron pressure
cannot be neglected and the total gas pressure needs to include both the proton
and electron contributions, namely $P_{g} = P_{i} + P_{e} = P_{i}(1+\beta)$,
where $\beta(z)\equiv T_e/T_i$ is the electron to proton temperature ratio and
is taken here as a free parameter. 
While it is well established that electron-ion equilibration
in the downstream might be only partial \cite[]{ghavamian01,ghavamian07}, 
in the presence of a precursor (either induced by the CRs or by the NRF), 
also upstream of the shock the level of equilibration becomes an unknown.

We note that the electron-ion equilibration level can be different between
upstream and downstream because the plasma conditions in these two regions are
different. Hence we distinguish between upstream and downstream using two
separate parameters, $\beta_{\rm up}$ and $\beta_{\rm down}$, respectively.
On the right side of Eq.~(\ref{eq:rh3}) we include the heating produced by the
compression of CR acting on the plasma plus the heating due to the damping of
magnetic waves.

The computation strategy we adopt is as follows: we first solve
Eqs.~(\ref{eq:rh1})-(\ref{eq:rh3}) for the upstream dynamics starting from the
boundary conditions at upstream infinity, i.e. $\rho_i(-\infty)=\rho_{i,0}$ and
$u_i(-\infty)= V_{\rm sh}$. In order to determine the conditions immediately
downstream of the subshock we use the jump conditions derived from the
conservation equations:
\begin{equation} \label{eq:JC1}
 \left[\rho_i u_{i} + \mu_N  \right]_1^2 = 0 \,,
\end{equation}
\begin{equation} \label{eq:JC2} 
 \left[ \rho_i u_{i}^{2} + P_{g} + P_{w} \right]_1^2 = 0 \,,
\end{equation}
\begin{equation} \label{eq:JC3}
 \left[ \frac{1}{2} \rho_i u_{i}^{3} + 
  \frac{\gamma_{g} P_{g} u_{i}}{\gamma_{g}-1} + F_w \right]_1^2 = 0 \,,
\end{equation}
where the brackets indicate the difference between quantities upstream and
downstream of the shock ($[X]_1^2 = X_2-X_1$). The terms describing neutrals and
CRs are absent because of the continuity of $f_{N}$ and $f$ across the
subshock. Finally we solve the conservation equations,
Eqs.~(\ref{eq:rh1})-(\ref{eq:rh3}) in the downstream region, using the values of
the quantities at position 2 as boundary conditions.

\begin{figure}
\begin{center}
{\includegraphics[width=0.7\linewidth]{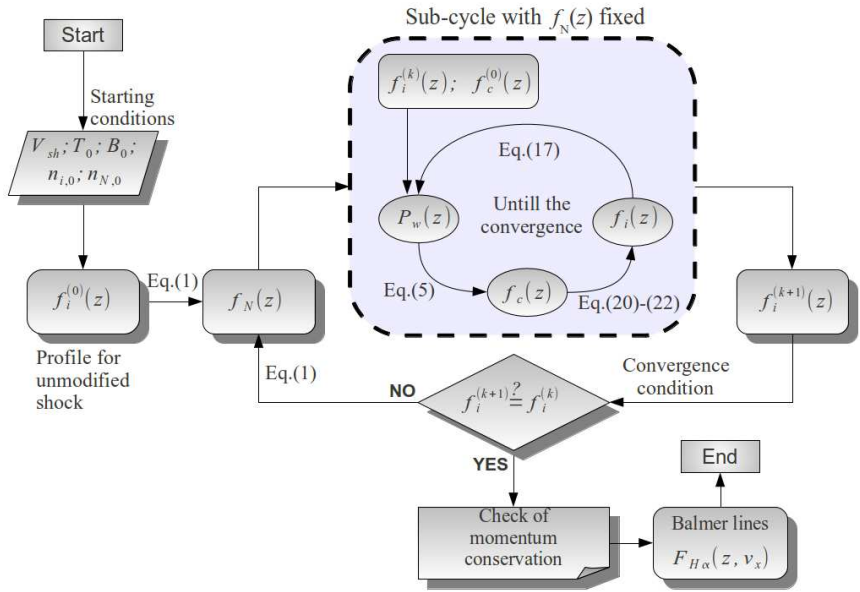}}
\caption{Flow chart of the iterative method used to find the solution.}
\label{fig:diagram}
\end{center}
\end{figure}

\subsection{Iterative method for the solution of the non-linear problem} 
\label{sec:algorithm} 

In order to solve the set of non-linear equations involving neutrals, ions, CRs
and magnetic field, we adopt an iterative method that we illustrate in
Fig.~(\ref{fig:diagram}) in the form of a flow diagram. The input quantities are
the values of the shock velocity and all environmental quantities at upstream
infinity, where the distribution function of neutrals is assumed to be
Maxwellian at the same temperature as that of ions. The method proceeds through
two nested cycles of iteration: in the first cycle, we start with a distribution
function of ions that corresponds to an unperturbed distribution that one would
obtain in the absence of both neutrals and CRs and we calculate the neutral
distribution function $f_{N}(z)$ from Eq.~(\ref{eq:vlasov}). With the given
$f_{N}(z)$ we iterate to obtain the ion distribution, the CR distribution and
the wave energy density. The outer loop of iteration allows us to update the
neutral distribution function using the newly available ion and CR distribution
functions. The iteration ends when the predefined accuracy is achieved. At this
point the distribution functions are used to calculate the structure of the
Balmer line, following the method illustrated in Paper II. 

The typical computation time required to complete the calculations for a given
set of environment parameters is of the order of one hour on a standard laptop.
The most time-consuming part is the solution of Eq.~(\ref{eq:vlasov}) which
increases linearly with the number of partial functions, $f_{N}^{(k)}$, required
for a correct description of the neutral distribution function. In particular
the number of $f_{N}^{(k)}$ increases linearly with the ratio between the
precursor length and the CE interaction length.

\section{Results} 
\label{sec:results}

The formalism presented above represents the first theory of particle
acceleration at collisionless shocks in the presence of neutral atoms and allows
us to calculate the shock structure, the spectrum of accelerated particles and
the Balmer emission from the shock region. In the subsections below we describe
the main results concerning these three topics. 

\subsection{Shock structure and spectra of accelerated particles}

\begin{figure}
\begin{center}
{\includegraphics[width=0.6\linewidth]{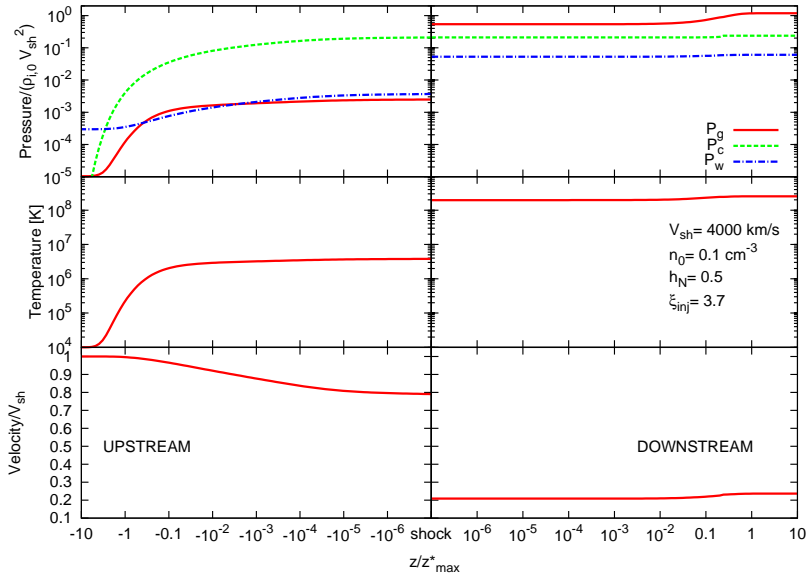}}
\caption{Example of spatial profiles of relevant quantities for a shock
propagating in partially neutral plasma. Top panel: pressure of ions, CRs and
magnetic field. Middle panel: ion temperature. Bottom panel: ion velocity.
Quantities are shown in the shock frame with the subshock located at the origin.
The $z$ axis is shown in logarithmic scale both upstream and downstream and $z^*_{\rm
max}= D(p_{\rm max})/V_{\rm sh}$ is the diffusion length of particles with the
maximum momentum. These profiles are computed for: $V_{\rm sh} =
4,000\,\rm km\,s^{-1}$, $T_0=10^4$~K, $B_0= 10\,\mu$G, $\rho_{i,0}=\rho_{N,0}=
0.05\,\rm cm^{-3}$. The injection parameter for CRs is $\xi_{\rm inj}= 3.7$,
which corresponds to an acceleration efficiency $\sim 20\%$.
}\label{fig:shock}
\end{center}
\end{figure}

We adopt a test case consisting of a shock moving at $V_{\rm sh} = 4,000\,\rm
km\,s^{-1}$ in a medium with temperature $T_0=10^4$~K, magnetic field $B_0=
10\,\mu$G, and densities $\rho_{i,0}=\rho_{N,0}= 0.05\,\rm cm^{-3}$ (the neutral
fraction is 50\%). CRs are injected with an efficiency of $\sim 20\%$ (in fact
we fix the injection parameter $\xi= 3.7$) and the efficiency of Turbulent
Heating (TH hereafter) is fixed as $\eta_{\rm TH}=0.5$ (see Eq.~\ref{eq:wave_tr2}). 
The shock structure is
illustrated in Fig.~\ref{fig:shock}: the top panel shows the pressure of gas
(continuous solid line), magnetic field (dotted line) and CRs (dashed line) as
functions of $z$, plotted here in logarithmic scale for both the upstream and
downstream
regions. All pressure components are normalized to the ram pressure
$\rho_{i,0}V_{\rm sh}^{2}$ at upstream infinity. Clearly the CR pressure is
continuous across the shock front. The self-generated magnetic field acquires a
pressure that is of the same order of magnitude of the thermal pressure, which
implies that its dynamical effect is non-negligible, as already discussed in
\cite{Caprioli08,Caprioli09}. 

The background ions are substantially heated in the precursor due to TH, as it
is shown by the increase in the gas temperature on a spatial scale
comparable with $z^*_{\max} \equiv z^*(p_{\max})\equiv D(p_{\max})/V_{sh}$,
i.e., the diffusion length of particles at $p_{\max}$ (mid panel of
Fig.~\ref{fig:shock}).
The upstream profile of the fluid velocity, showed in the bottom panel of
Fig.~\ref{fig:shock} as normalized to $V_{\rm sh}$, is a distinctive signature
of the CR-induced precursor.

Finally, one last thing to notice in this Figure is the increase of both the
gas pressure and velocity at some distance from the shock in the downstream.
This is a result of progressive ionization of fast cold neutrals, and indeed
occurs on scales which are of the order of the ionization length. We will see
that the situation is different, with the plasma velocity decreasing, rather than 
increasing, towards downstream infinity, when the shock velocity is lower and
the NRF starts to play an important role (see Fig.~\ref{fig:slope_fN} below).
Spatial variations of the plasma velocity are important for determining the
accelerated particle spectrum at the shock, whose slope at each given energy
always reflects the mean compression ratio experienced by particles of that
energy while moving away from the shock.

The spectrum of accelerated particles at different locations upstream and
downstream of the shock is shown in Fig.~\ref{fig:f(p,x)} for a neutral fraction
$h_{N}=0.5$. The solid lines refer to the downstream section of the plasma, the
thick dot-dashed line is the spectrum at the subshock location, the dashed lines
are the spectra at different locations upstream of the shock. 

One can see that, moving away from the shock in the upstream direction, the spectrum is
more and more depleted of the lowest energy particles: at any position $z$,
only particles energetic enough to have a diffusion length $D(p)/V_{\rm sh}
\gtrsim z$ are present. The transport of low-energy particles in the downstream,
is rather dominated by advection, and the spectrum is never truncated: 
at downstream infinity, the low-energy spectrum has the same slope as at the
shock, but with a higher normalization.
This is due to the compression that occurs around $0.1<z/z_{\max}^*<1$, where
ionization takes place.  At the highest energies, instead, diffusion dominates
over advection and the CR spectrum far downstream is expected (and found) to be
exactly the same as at the shock location. As a result, at large distances
downstream of the shock the spectrum turns out 
to be slightly steeper than at the shock.

As already illustrated in Paper I in the test-particle limit, the effect
of the NRF is to create an upstream precursor that steepens the spectra of
accelerated particles. For this reason the spectra of accelerated particles are
less concave than they would be in the presence of the CR reaction alone. This
result is qualitatively general, though the strength of the effect is
quantitatively dependent upon the specific environment that the shock propagates
into, as discussed below.

\begin{figure}
\begin{center}
{\includegraphics[width=0.5\linewidth]{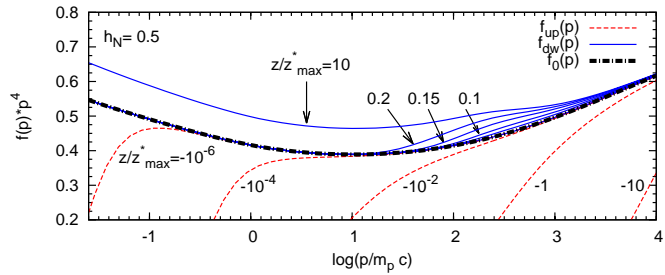}}
\caption{CR momentum spectrum multiplied by $p^4$ for different distances from
the sub-shock: the dash-dotted (black) line is the spectrum at the sub-shock position,
the dashed (red) lines show the spectrum in the upstream and solid (blue) lines
show the spectrum downstream. The position $z$ is indicated for each curve in
units of $z_{\max}^*$.}
\label{fig:f(p,x)}
\end{center}
\end{figure}

In Fig.~\ref{fig:slope_fN} we show the normalized plasma velocity 
as a function of the distance from the shock for different values of the
neutral fraction (left panel), and the slope of the spectrum of accelerated
particles as a function of momentum (right panel). All the calculations refer to
a shock speed $V_{\rm sh}=2,000\,\rm km\,s^{-1}$, a total gas density at upstream
infinity $n_{0}=0.1\,\rm cm^{-3}$, an unperturbed magnetic field strength
$B_{0}=10\,\mu$G and an injection parameter $\xi_{\rm inj}=3.7$. The solid lines
refer to the case with fully ionized plasma ($h_{N}=0$).
In this case the shock is appreciably modified by the CR pressure, the acceleration
efficiency being $\sim 30\%$, and the spectrum is quite concave. The
spectrum is steeper than $p^{-4}$ only for energies below $\sim 10$ GeV. 

Due to the relatively-low shock velocity, the effects of CE and ionization are
quite relevant; therefore, we expect the results to change appreciably for non
vanishing values of $h_{N}$, due to the NRF. This is illustrated by the
long-dashed, dashed and dotted lines in Fig.~\ref{fig:slope_fN}. For $h_{N}=0.2$
(long-dashed line) and $h_{N}=0.5$ (dashed line) the formation of the precursor
upstream is dominated by the dynamical reaction of CRs on the plasma, although
the heating of the upstream due to the NRF is very evident. In these two cases
the acceleration efficiency is $\sim 20\%$ (for $h_{N}=0.2$) and $\sim 15\%$
(for $h_{N}=0.5$). The spectra in the same cases are correspondingly less
modified by the CR pressure (less concave) and, on average, somewhat steeper: for
$h_{N}=0.2$ the spectrum is steeper than $p^{-4}$ for energies  $\lesssim 100$
GeV, while for $h_{N}=0.5$ it remains steeper than $p^{-4}$ even at $E\sim
10^{4}$ GeV. 

The case with large neutral fraction, $h_{N}=0.8$ is rather extreme: the
precursor becomes very short, as it becomes dominated by the NRF, and the
acceleration efficiency is still $\sim 15\%$. The spectrum is quite steeper than
$p^{-4}$ at all energies, so that the bulk of the pressure is carried by
relatively low energy particles. This implies that the CR pressure can
slow down the incoming plasma only very close to the subshock
(see left panel in Fig.~\ref{fig:slope_fN}). 

\begin{figure}
\begin{center}
{\includegraphics[angle=0,width=0.45\linewidth]{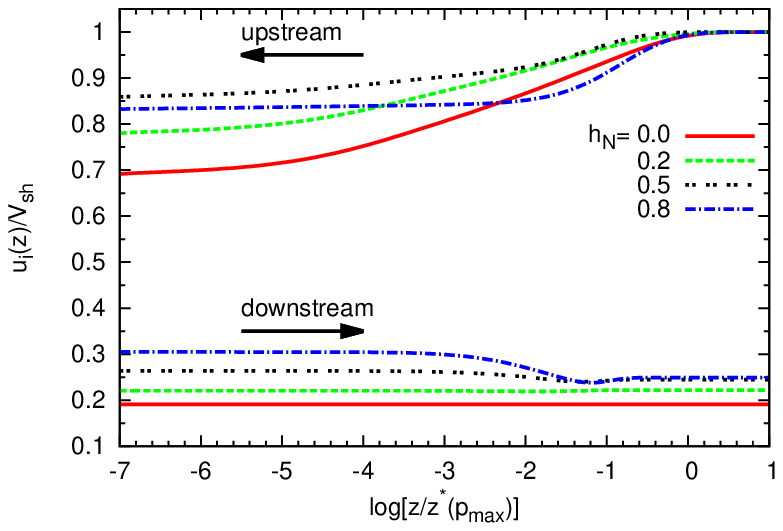}}
{\includegraphics[angle=0,width=0.45\linewidth]{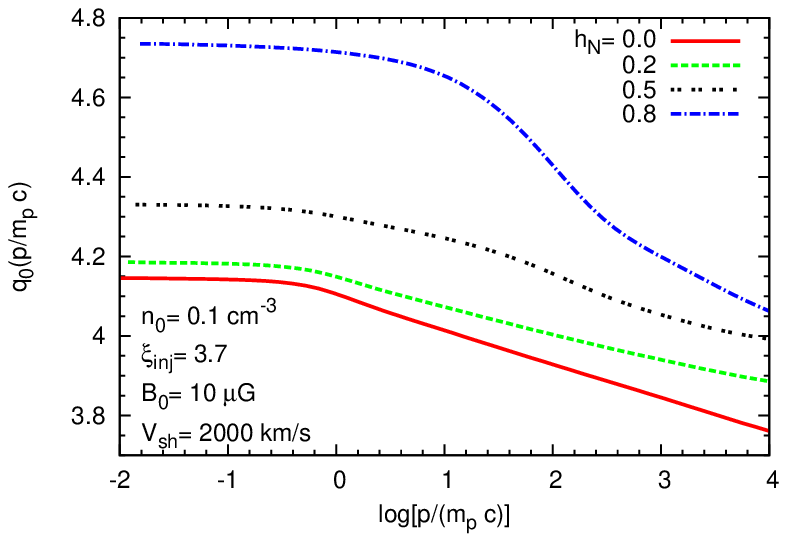}}
\caption{{\it Left panel}: 
Velocity profile of the upstream and downstream
fluids normalized to the shock velocity, as a function of the  distance from the
shock. The shock is located on the left side and the (absolute) distance is in unit of
$z^*(p_{\rm max})$. 
The profiles are shown for a given shock speed $V_{\rm sh}=2,000$km s$^{-1}$
and for different neutral fractions $h_{N}$ from 0 up to 80\% as in the legend.
Other parameters are $\xi_{\rm inj}=3.7$, $B_0= 10\,\mu$G and $n_0=0.1\,\rm
cm^{-3}$. {\it Right panel}: Slope of the spectrum of accelerated particles for
the same shock configurations as in the left panel.
}
\label{fig:slope_fN}
\end{center}
\end{figure}

\begin{figure}
\begin{center}
{\includegraphics[angle=0,width=0.45\linewidth]{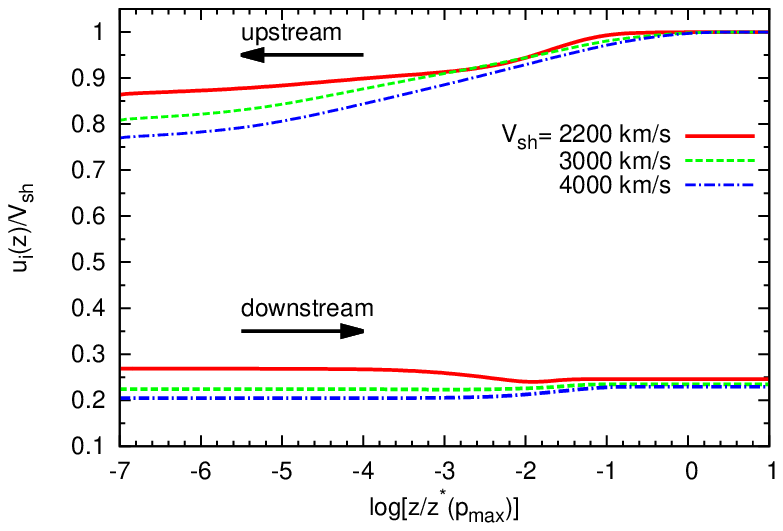}}
{\includegraphics[angle=0,width=0.45\linewidth]{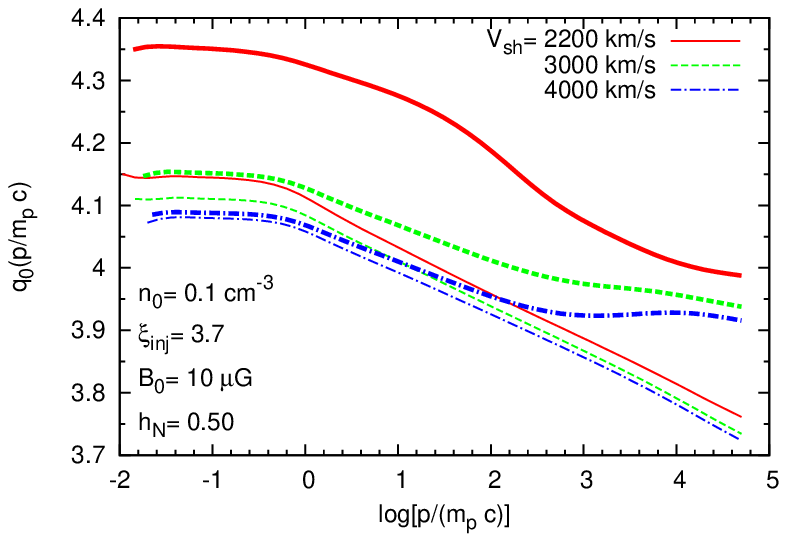}}
\caption{
As in Fig.~\ref{fig:slope_fN}, but for a fixed neutral fraction $h_{N}=0.5$ and
different shock velocities, as indicated. For comparison, in the right panel the
thin lines show the slope in the absence of neutrals.
}
\label{fig:slope_Vsh}
\end{center}
\end{figure}

As discussed in Paper I, the effects of the NRF are most important for shock
speeds $V_{\rm sh}\leq 3,000\,\rm km\,s^{-1}$; for larger shock speeds, the
typical relative velocity between neutrals and ions is such that the
cross section for CE is so small that atoms get ionized before suffering
a CE reaction. As a consequence, the NRF is suppressed. It is useful to
explore the dependence of the spectrum on the shock velocity in a case in which the
neutral fraction is $h_{N}=0.5$. In the left panel Fig.~\ref{fig:slope_Vsh} we
plot the normalized plasma velocity as a function of distance
from the shock for shock velocities $V_{\rm sh}=2,200\,\rm km\,s^{-1}$ (solid line),
$V_{\rm sh}=3,000\,\rm km\,s^{-1}$ (dashed line) and $V_{\rm sh}=4,000\,\rm km\,s^{-1}$ (dotted line). The other
parameters are the same as in Fig.~\ref{fig:slope_fN}. The right panel shows
 the spectral slope for the same values of the shock velocity. For
comparison, the thick lines refer to $h_{N}=0.5$, while the thin lines of the same
type refer to completely ionized gas ($h_{N}=0$).

The left panel qualitatively confirms our expectations based on the
test-particle results: when increasing the shock speed from $2,200\,\rm
km\,s^{-1}$ to $4,000\,\rm km\,s^{-1}$, the effect of neutrals becomes smaller,
namely the shock becomes increasingly more modified by the reaction of CRs. This
is even more evident in the right panel of Fig.~\ref{fig:slope_Vsh}: for $V_{\rm
sh}=2,200\,\rm km\,s^{-1}$ a neutral fraction of 50\% is sufficient to make the
spectrum of accelerated particles steeper than $p^{-4}$, at all energies (thick
solid line). With the same shock speed, but for a completely ionized plasma, 
the spectrum is instead harder than $p^{-4}$ above $\sim 10$GeV ($h_{N}=0$, thin
solid line). 

\subsection{Balmer emission}
\label{sec:balmer}

The Balmer line is affected by the presence of CRs in at least two ways. First
of all, the width of the broad Balmer line is expected to be narrower if
efficient CR acceleration takes place. This is a straightforward  consequence of
energy conservation: energy is channelled into accelerated particles, therefore
the plasma temperature downstream is lower.
Moreover, the formation of a CR-induced precursor leads to a non-negligible
heating upstream of the subshock, so that the narrow Balmer line may
become broader. In this section we investigate these possibilities in a more
quantitative way, by using the formalism introduced in Paper II. The different
components of the line are identified as discussed in Paper II. The global shape
of the Balmer line is fitted with three Gaussian components: a narrow line, an
intermediate line (reflecting directly the effect of the NRF) and a broad line.

\begin{figure}
\begin{center}
{\includegraphics[angle=0,width=0.5\linewidth]{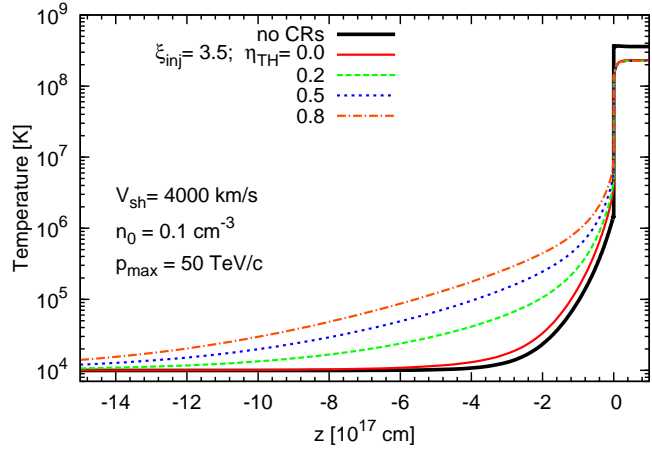}}
\caption{Temperature of the plasma in the shock region for $V_{\rm
sh}=4,000\,\rm km\,s^{-1}$, $n_0= 0.1\,\rm cm^{-3}$, $h_N= 0.5$ and
$B_0=10\,\mu$G. When CR acceleration is turned on, we assume $\xi_{\rm inj}=3.5$
and $p_{\rm max}=50$~TeV/c. The thick solid (black) line refers to the case with no CRs. The thin (red)
solid line refers to the case with CRs but no TH. Different values of $\eta_{\rm
TH}$ are as indicated.}
\label{fig:Temp_prec}
\end{center}
\end{figure}

In Fig.~\ref{fig:Temp_prec} we show the temperature throughout the shock region
for different values of the efficiency of TH, $\eta_{\rm TH}$, for a shock
propagating at $4,000\,\rm km\,s^{-1}$ in a medium with density $0.1\,\rm cm^{-3}$.
The case without CR acceleration is shown as a thick solid (black) line. In this
case the precursor is solely due to the NRF and is limited to the region very
close to the subshock. 

When the acceleration process is turned on, several phenomena
become visible. In the absence of TH ($\eta_{\rm TH}=0$, thin solid line),
the upstream fluid is mildly heated by the adiabatic compression due to the CR-induced precursor only. 
Nevertheless, the temperature of the downstream plasma
drops by almost a factor 2 as a consequence of the energy channelled into the
accelerated particles. Increasing the value of $\eta_{\rm TH}$, the heating in
the precursor becomes more and more marked, and it extends over larger and larger distances. 
However, the  downstream temperature  is not appreciably affected by a change 
in the efficiency of TH.

\begin{figure}
\begin{center}
{\includegraphics[width=0.45\linewidth]{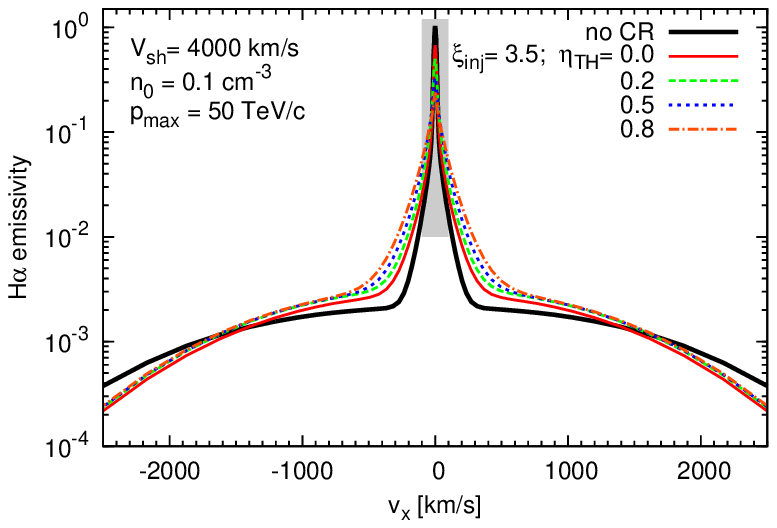}}
{\includegraphics[width=0.45\linewidth]{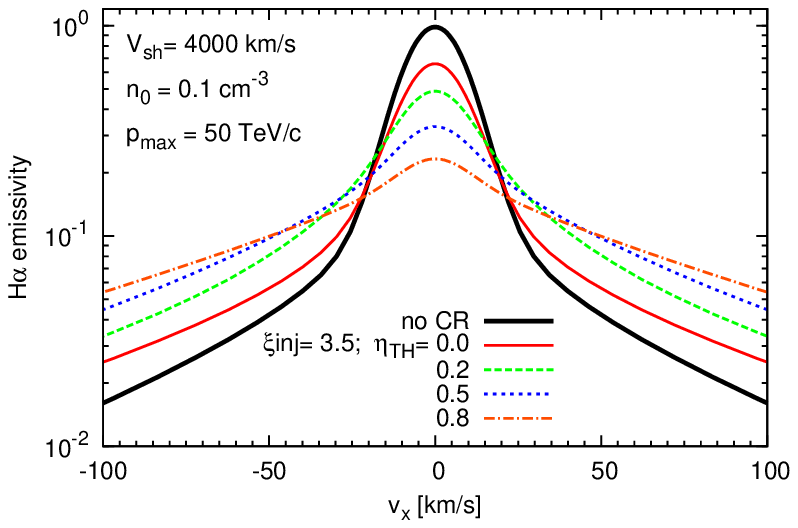}}
\caption{Volume-integrated line profile of Balmer emission for the same cases
shown in Fig.~\ref{fig:Temp_prec}. The right panel shows a zoom on the narrow
line region, i.e. the shadow box in the left panel. The thick solid line shows
the case without CRs, while different thin lines are calculated at fixed $\xi_{\rm
inj}= 3.5$ for different values of the TH efficiency.}
\label{fig:Balmer1}
\end{center}
\end{figure}

The shape of the global Balmer line emission is plotted in the left panel of
Fig.~\ref{fig:Balmer1}, while the right panel shows a zoom in the region of the
narrow Balmer line. The curves refer to the same cases as in
Fig.~\ref{fig:Temp_prec}, labelled as indicated. The width of the broad
component of the Balmer line is appreciably reduced when CR acceleration is
turned on, but the results are not sensitive to the amount of TH. On the other
hand, as one can notice in the right panel, the narrow Balmer line
becomes broader when CR acceleration is efficient; the effect is more
pronounced when non-negligible TH is taken into account. 
Hence, the distribution function of neutrals becomes broader mainly because of the
scattering with a warmer ion distribution in the far precursor (i.e., where
TH is more effective), rather than because of the NRF, which operates only within a
few CE interaction lengths from the subshock. 

\begin{figure}
\begin{center}
{\includegraphics[width=0.45\linewidth]{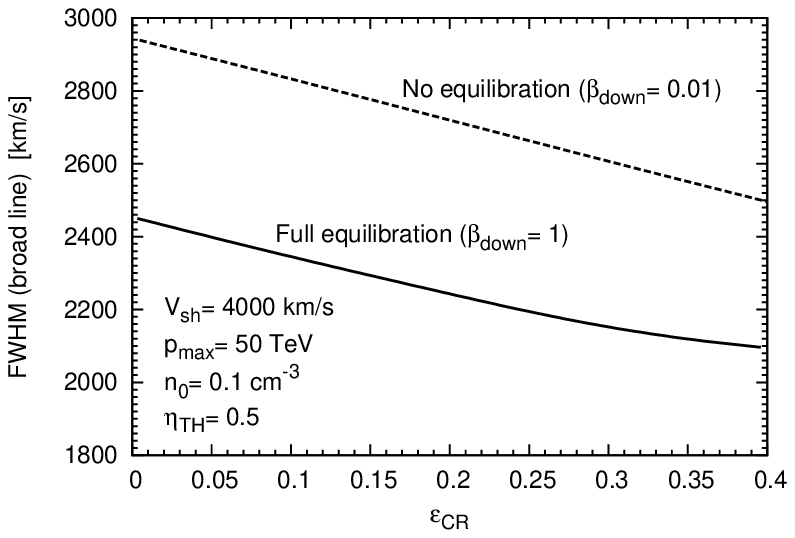}}
{\includegraphics[width=0.45\linewidth]{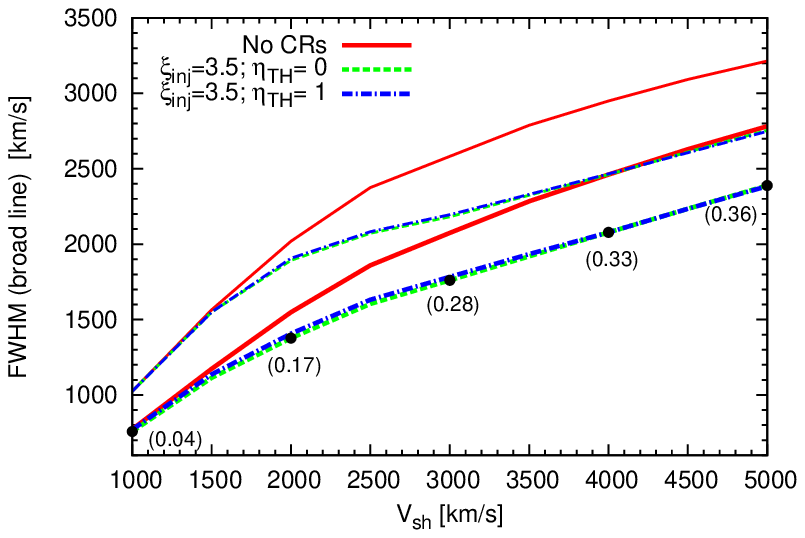}}
\caption{FWHM of the broad Balmer line as a function of the CR acceleration
efficiency $\epsilon_{\rm CR}$ (left panel) and as a function of the shock
velocity $V_{\rm sh}$ (right panel). The upper set of curves (thin lines)
correspond to the case $\beta_{\rm down}=0.01$, while the lower curves (thick
lines) correspond to $\beta_{\rm down}=1$. In the right panel, the values near
the points refer to the associated CR acceleration efficiency.}
\label{fig:Balmer2}
\end{center}
\end{figure}

In Fig.~\ref{fig:Balmer2} we show the FWHM of the broad Balmer line as a
function of the CR acceleration efficiency $\epsilon_{\rm CR}$ (left panel) and
as a function of the shock velocity $V_{\rm sh}$ (right panel). In the left
panel, the two curves illustrate the change in the FWHM obtained by assuming
full or partial temperature equilibration between electrons and protons in the downstream 
 ($\beta_{\rm down}=1$ and $\beta_{\rm down}=0.01$ , with $\beta_{\rm down}$ the
ratio between electron and ion temperature), for $V_{\rm sh}=4,000\,\rm
km\,s^{-1}$ and $n_0=0.1\,\rm cm^{-3}$. 

The FWHM decreases when the energy density in CRs increases, 
confirming the very important notion that the width of the
broad Balmer line can be used to measure the CR acceleration efficiency.
There is however an important caveat to keep in mind: for a
given value of $\epsilon_{\rm CR}$, the uncertainty in the electron-proton equilibration
translates into a spread in the FWHM of $\sim 500\,\rm km\,s^{-1}$.
Hence, unequivocal evidence of efficient CR acceleration is likely to be achievable only if the
measured FWHM is below the line corresponding to full equilibration. Even in
this case the value of $\epsilon_{\rm CR}$ is likely to be quite uncertain
because of the unknown level of equilibration. For instance, if a FWHM of $2,200\,\rm km\,s^{-1}$ were measured
for a shock with $V_{\rm sh}=4,000\,\rm km\,s^{-1}$, the result could be
interpreted as a measurement of $\epsilon_{\rm CR}\sim 0.25$ in the case of full
equilibration. In the more general case of partial equilibration, the above
value of $\epsilon_{CR}$ can be taken as a lower limit. On the other hand, for
the same shock velocity, if one measured a FWHM of $2,600\,\rm km\,s^{-1}$, it
would be hardly possible to say anything at all about the efficiency of CR
acceleration, without knowing the level of electron-proton thermalization.
In order to have an independent estimate of $\beta$, it would be 
very helpful to pair optical observations with the emission of SNR shocks in the 
(thermal) X-rays as well.

In the right panel of Fig.~\ref{fig:Balmer2} we fix the injection parameter
$\xi_{\rm inj}=3.5$ and we calculate the FWHM of the broad Balmer line as a
function of the shock velocity. The two thin (red) lines refer to the cases with
no CR acceleration with $\beta_{\rm down}=0.01$ (upper curve) and
$\beta_{\rm down}=1$ (upper curve). The dashed ($\eta_{\rm TH}=0$) and
dash-dotted lines ($\eta_{\rm TH}=1$) refer to the cases with CR acceleration
(efficiencies in parentheses), again for $\beta_{\rm down}=0.01$ (upper curves)
and $\beta_{\rm down}=1$ (lower curves). From the plot it is clear that the TH
does not affect appreciably the width of the broad line.
As pointed out above, if the electron-proton thermalization is inefficient, it
may be difficult to draw any strong conclusion about the CR acceleration
efficiency unless additional information is available to constrain the value of
$\beta_{\rm down}$. Nevertheless, measuring a FWHM of the broad line below the
solid thick line allows one to put a lower limit on the CR acceleration
efficiency.

\begin{figure}
\begin{center}
{\includegraphics[angle=0,width=0.45\linewidth]{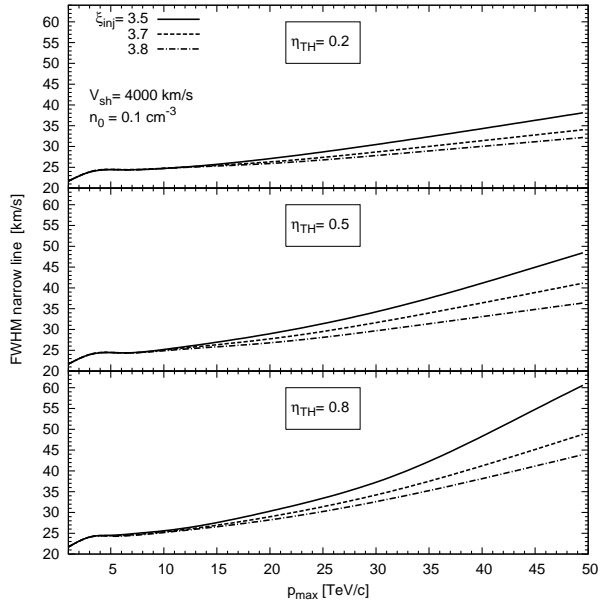}}
\caption{FWHM of the narrow line as a function of the maximum momentum of
accelerated protons. The three panels refer to $\eta_{\rm TH}=$0.2, 0.5 and 0.8.
Each panel shows three lines calculated assuming different injection parameter,
$\xi_{\rm inj}=$3.5, 3.7 and 3.8 which approximately correspond to
$\epsilon_{\rm CR}=$ 0.4, 0.2 and 0.1.}
\label{fig:narrow}
\end{center}
\end{figure}

Additional discrimination power might derive from a combined analysis of the
broad and narrow Balmer line components. The FWHM of the narrow Balmer line as a
function of the maximum momentum $p_{\rm max}$ is plotted in
Fig.~\ref{fig:narrow} for $\eta_{\rm TH}=0.2$ (top left panel), $\eta_{\rm
TH}=0.5$ (top right panel) and $\eta_{\rm TH}=0.8$ (bottom panel). The three
curves in each panel are obtained for $\xi_{\rm inj}=3.5,~3.7,~3.8$. The shock
velocity and total density are set to $V_{\rm sh}=4,000\,\rm km\,s^{-1}$ and
$n_{0}=0.1\,\rm cm^{-3}$, which approximately correspond to $\epsilon_{\rm CR}=$
0.4, 0.2 and 0.1 (weakly dependent upon $\eta_{\rm TH}$) for the three values of
$\xi_{\rm inj}$. The maximum momentum, $p_{\rm max}$ determines the spatial
extent of the CR-induced precursor. Larger values of $p_{\rm max}$ imply that
there is more time (space) for depositing heat in the upstream,
and the width of the narrow Balmer line broadens correspondingly. 
The effect becomes more pronounced for larger values
of the parameter $\eta_{\rm TH}$. It is worth noticing that the behavior of the
curves in Fig.~\ref{fig:narrow} is the same in all cases and reflects the slight
broadening of the narrow Balmer line as due to the NRF. In fact, one can estimate
the momentum at which the diffusion length equals the size of the precursor
induced by the NRF, which is of order $\lambda_{\rm NRF}\sim 10^{16}$~cm for the
parameters used here. By imposing $D(p_{\rm max})/V_{\rm sh}={p_{\rm max}c^{2}}/{3e\,\delta
B}=\lambda_{\rm NRF}$ one gets $p_{\rm max}\sim 5-10$~TeV/c (from our
calculations the amplified magnetic field close to the subshock is $\delta
B\sim\,50\mu$G). This interpretation of the modest increase in the FWHM for low
values of $p_{\rm max}$ is further confirmed by the fact that the width is very
weakly dependent on both $\xi_{\rm inj}$ and $\eta_{\rm TH}$ for $p_{\rm
max}\lesssim 10$~TeV/c.

\begin{figure}
\begin{center}
{\includegraphics[width=0.45\linewidth]{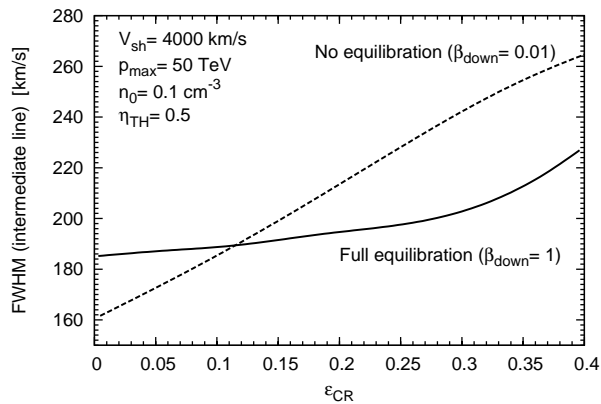}}
\caption{FWHM of the intermediate Balmer line as a function of the CR
acceleration efficiency, $\epsilon_{\rm CR}$, for a shock speed $V_{\rm
sh}=4,000\,\rm km\,s^{-1}$ $\eta_{\rm TH}=0.5$ and $p_{\rm max}=50$~TeV. The solid
and dashed lines refer to the cases $\beta_{\rm down}=1$ and
$\beta_{\rm down}=0.01$
respectively.}
\label{fig:intermediate}
\end{center}
\end{figure}

For completeness, Figure~\ref{fig:intermediate} shows the width of the
intermediate component of the Balmer line as a function of the CR acceleration
efficiency $\epsilon_{\rm CR}$, for $V_{\rm sh}=4,000\,\rm km\,s^{-1}$,
$\eta_{\rm TH}=0.5$ and neutral fraction $h_{N}=0.5$. The two curves refer again
to $\beta_{\rm down}=0.01$ (upper curve) and $\beta_{\rm down}=1$ (lower
curve). 
In the upstream, energy is assumed to be deposited only in the
protons, which therefore become warmer than the electrons.
We assume that they do not equilibrate very effectively since the Coulomb
collision timescale is typically longer than the advection one.

The reason why the width of the intermediate Balmer line depends on the CR 
acceleration efficiency is that when $\epsilon_{\rm CR}$ increases, 
the amount of TH close to the subshock increases as well,
leading to a broader distribution of neutrals in the region affected by the NRF.
The dependence of the width of the intermediate line on $\beta_{\rm down}$ 
lies in the fact that the NRF carries upstream the information about 
the ion temperature behind the shock.
We further notice that the width of the intermediate component may also
depend on the electron-ion equilibration upstream, $\beta_{\rm up}$. In
Figure~\ref{fig:intermediate} we assume no equilibration upstream, namely while
ions are heated by NRF and TH, electrons always have $T= 10^4$ K. Hence one
has to be careful in using the intermediate line to infer shock properties, 
because this line is a function of several parameters.

\section{Summary and Conclusions}
\label{sec:conclusion}

In this paper we have presented the first formulation of the non-linear
theory of diffusive shock acceleration at SNR shocks propagating in a partially
ionized medium. At the same time, we have put forward the first fully
self-consistent calculation of how the shape of Balmer ${\rm H}_\alpha$ line is
affected by the presence of accelerated particles. This development is crucial
if one wants to use the width of the various components of the Balmer
line to infer information on the efficiency of CR acceleration in SNRs.

The three most important physical points discussed here in a quantitative way
are the following: 1) neutrals produced in CE events with hot ions downstream
can return upstream and deposit energy and momentum there. This phenomenon, that
we refer to as the neutral return flux (NRF), has an important dynamical effect on the
structure of collisionless shocks as well as on the spectrum of accelerated
particles (see also Paper I). 2) The structure of a collisionless shock is
affected in a non-trivial way by the combined action of neutrals and CRs. 3)
Also the shape of the Balmer ${\rm H}_\alpha$ line reflects the effects of the
two phenomena mentioned above.

The NRF was not predicted in most of the previous calculations of the structure of SNR
shocks, mainly because of their intrinsic limitation to fluid approaches 
(that cannot account for this kinetic effect) or
restriction to the downstream plasma alone. Both CRs and the NRF produce a
precursor upstream of the shock: the CR-induced precursor develops on a scale of
the diffusion length of particles at the maximum energy $\sim D(p_{\rm
max})/V_{\rm sh}$, while the neutral-induced precursor develops on a scale of a
few CE interaction lengths. Both precursors cause heating of the ion plasma
upstream. The heating associated with the NRF is mediated by CE and ionization
upstream. The CR-induced heating is most effective only in the presence of
turbulent (namely non adiabatic) heating.

For typical values of the parameters relevant for SNRs, the spatial extent of
the neutral-induced precursor is such as to overlap with the diffusion range of
particles with $p\lesssim 5-10$~TeV/c. In this energy region the spectrum of
accelerated particles is steepened by the dynamical effect of the NRF, as
already found in Paper I, in the context of test-particle calculations. The
calculations presented here are a generalization of previous work on NLDSA
\cite[]{AmatoBlasi05,AmatoBlasi06,Caprioli09} and on the dynamical effects of
neutrals on the structure of collisionless shocks \cite[]{paperI}. The
technique adopted for the solution of the non-linear equations that describe the
whole problem is basically an iterative method in two stages: in a first stage,
we calculated iteratively the distribution of neutrals and ions (see
Sections~\ref{sec:neutrals}, \ref{sec:dynamics} and Paper I). The distribution
function at a generic location is not Maxwellian therefore the calculation of
the neutral distribution is kinetic, while ions (protons and electrons) are
treated as fluids. In a second stage, the updated profile of neutrals and ions
is used to calculate the spatial distribution and spectrum of accelerated
particles and waves (Sections~\ref{sec:CRspectrum} and \ref{sec:waves}). The
procedure is repeated until convergence is reached. 

On average, as expected, the presence of neutrals leads to less concave and
steeper spectra of accelerated particles than in a completely ionized medium.
These effects are quite substantial for shocks with $V_{\rm sh}\lesssim
3,000\,\rm km\,s^{-1}$. For faster shocks, the NRF becomes dynamically less
important because neutrals are ionized downstream before they suffer a CE
reaction. In these cases the shock modification is mainly induced by CRs. 

Besides the general interest in a theory of NLDSA in the presence of neutrals,
we think that the present work is especially important in terms of providing an
essential tool to interpret the shape of the ${\rm H}_\alpha$ line and translate
it into quantitative information on the presence of CRs in SNR shocks. The
widths of the narrow and broad component of the Balmer line in SNR shocks have
long been known to bear information on the temperature of the ions upstream and
downstream of the shock respectively \cite[see][and the recent high quality results of \cite{SN1006}]{raymond78,vanadel, smith91, ghavamian01, heng07}. Both quantities
are sensitive, in turn, to the CR acceleration efficiency: if CRs are
efficiently accelerated, the plasma upstream is heated by the CR-induced
precursor and the downstream plasma is heated less than in the absence of CRs
because part of the ram pressure is channelled into CR acceleration. In terms of
Balmer line, one expects the narrow component to be somewhat broader and the
broad component to be somewhat narrower whenever CR acceleration is efficient. A
phenomenological approach to the problem was recently put forward by
\cite{raymond2011}, where however the CR spectrum and spatial distribution are
both assumed rather than calculated. As described in the present paper, the CR
properties are profoundly affected by the presence of neutrals. Moreover
\cite{raymond2011} do not treat the downstream plasma, therefore there is no NRF
in their calculation. 

In Fig.~\ref{fig:Temp_prec} we can see the temperature of the plasma calculated
self-consistently. Both the heating in the precursor (for different values of
the efficiency of TH) and the reduced temperature in the downstream (as a result
of CR acceleration) are clearly visible. 

These effects reflect in the shape of the Balmer line, which we calculated by
using the formalism introduced in Paper II. The width of the narrow Balmer line
is found to depend appreciably upon the level of TH and the value of maximum
momentum $p_{\rm max}$. The dependence on $p_{\rm max}$ can be understood,
because increasing $p_{\rm max}$ leads to larger CR precursors and therefore
neutrals incoming from upstream infinity have a longer time to experience CE and
adapt to the ion temperature. Larger TH reflects in an increased width of the
narrow Balmer line. 

The level of TH does not affect appreciably the plasma temperature downstream,
therefore the width of the broad Balmer line mainly reflects the CR acceleration
efficiency. However, the broad line is sensitive to the level of equilibration
between protons and electrons downstream: if electrons and protons reach thermal
equilibrium in a time short compared with CE (clearly this cannot happen through
collisions) then the ion temperature is lower, thereby mimicking the presence o
CRs. The estimate of the CR acceleration efficiency $\epsilon_{\rm CR}$ might be
less dependent upon this ambiguity if the width of the broad Balmer line is
quite small. These cases however, suggesting a very high acceleration
efficiency, are very difficult to obtain in the presence of TH and magnetic
reaction on the shock \cite[]{Caprioli08,Caprioli09}. Moreover these cases would
correspond to very concave spectra of accelerated particles and of the
associated gamma rays, that so far have never been observed. The 
observation of the structure of the Balmer line (intensity and width of its components)
might also help breaking the
degeneracy between CR acceleration efficiency and electron-proton equilibration.
By the same token, observation of an X-ray spectrum of thermal origin may
independently provide a measurement of the electron temperature. 

Finally, we stress that the existence of the NRF leads to the formation of an
intermediate component of the Balmer line, with a width that depends on the
temperature of protons downstream, on the electron-proton equilibration and on
the level of TH. The practical possibility to infer $\epsilon_{\rm CR}$ from
data on individual SNR will need to be assessed on the case-by-case basis. 

\section*{Aknowledgments}
We are grateful to the anonymous referee for several comments that helped us improve the paper.
This work was partially funded through grant ASI-INAF I/088/06/0 and PRIN INAF
2010. The research work of D.C. was partially supported by NSF grant
AST-0807381.

\end{document}